\documentclass[aps,prl,showpacs,twocolumn,superscriptaddress,letterpaper]{revtex4}

\usepackage{graphicx}% Include figure files
\usepackage{dcolumn}% Align table columns on decimal point
\usepackage{bm}% bold math
\usepackage{amsmath, amssymb}% 
\usepackage{epstopdf}
\usepackage{color}
\begin{document}

%\modulolinenumbers[5]
%\linenumbers

%\preprint{}

\title{Hammer events, neutrino energies, and nucleon-nucleon correlations}

\newcommand*{\ODU }{ Old Dominion University, Norfolk, VA 23529,
USA} 
\newcommand*{\ODUindex}{1}
\affiliation{\ODU } 
\newcommand*{\MIT }{Massachusets Institute of Technology, Cambridge,
  MA 02139, USA}
\newcommand*{\MITindex}{2}
\affiliation{\MIT} 
\newcommand*{\TAU }{Tel Aviv University, Tel Aviv 69978, Israel}
\newcommand*{\TAUindex}{3}
\affiliation{\TAU }

%%%%%%%%%%%%%%%%%%%% authors %%%%%%%%% 
  
\author{L.B.~Weinstein}
     \email[Contact Author \ ]{weinstein@odu.edu}
     \affiliation{\ODU}
\author{O. Hen}
     \affiliation{\MIT}
\author{Eli Piasetzky}
     \affiliation{\TAU}

\date{\today}

\begin{abstract}

  {\bf Background:} Accelerator-based neutrino oscillation
  measurements depend on observing a difference between the
  expected and measured rate of neutrino-nucleus interactions at
  different neutrino energies or
  different distances from the neutrino source. Neutrino-nucleus
  scattering cross sections are complicated and depend on the neutrino
  beam energy, the neutrino-nucleus interaction, and the structure of
  the nucleus.  Knowledge of the incident neutrino energy spectrum and
  neutrino-detector interactions are crucial for analyzing neutrino
  oscillation experiments. The ArgoNeut liquid
  Argon Time Projection Chamber (lArTPC) observed charged-current
  neutrino-argon scattering events with two protons back-to-back in
  the final state (``hammer'' events) which they associated with
  short-range correlated (SRC) nucleon-nucleon
  pairs.  The large volume MicroBoone lArTPC will measure far more of
  these unique events.

  {\bf Purpose:} Determine what we can learn about the incident
  neutrino energy spectrum and/or the structure of SRC from
  hammer events that will be measured in MicroBooNE.

  {\bf Methods:} We simulate hammer events using two models and the
  well known electron-nucleon scattering cross section. In the first
  model the neutrino (or electron) scatters from a moving proton,
  ejecting a $\pi^+$, and the $\pi^+$ is then absorbed on a moving
  deuteron-like $np$ pair.  In the second model the neutrino (or
  electron) scatters from a moving nucleon, exciting it to a $\Delta$
  or $N^*$, which then deexcites by interacting with a second nucleon:
  $\Delta N\rightarrow pp$.

  {\bf Results:} The pion production and reabsorption process results
  in two back-to-back protons each with momentum of about 500 MeV/c,
  very similar to that of the observed ArgoNeut events.  These
  distributions are insensitive to either the relative or
  center-of-mass momentum of the $np$ pair that absorbed the $\pi$.
  In this model, the incident neutrino energy can be reconstructed
  relatively accurately using the outgoing lepton.  The $\Delta
  p\rightarrow pp$ process results in two protons that are less
  similar to the observed events.

  {\bf Conclusions:} ArgoNeut hammer events can be described by a
  simple pion production and reabsorption model.  The hammer events
  that will be measured in MicroBooNE can be used to determine the
  incident neutrino energy but not to learn about SRC.  We suggest that this reaction
  channel could be used for neutrino oscillation experiments to
  complement other channels with higher statistics but different
  systematic uncertainties.
\end{abstract}

%\keywords{}

\maketitle

%\section{}
%\section{Introduction \label{intro}}
%

%\setpagewiselinenumbers
%\linenumbers

\noindent{\bf Introduction:} 
Neutrino scattering from nuclei can be used to learn about both the energy
distribution of the incident neutrino beam and the
neutrino-nucleus interaction.  The neutrino beam energy distribution
is necessary to interpret the results of neutrino oscillation
experiments \cite{mosel16}.  The neutrino-nucleus interaction can be used
to understand the neutrino-detector interactions which complicate
interpretation of neutrino experiments or to learn more about the
structure of nuclei.

With large-volume liquid-argon time projection chambers
(lArTPCs), neutrino scattering experiments can measure all of the
charged particles emitted in an interaction, helping  disentangle
the effects of the neutrino energy distribution, the neutrino-nucleus
interaction, and nuclear structure.  

An important class of nuclear structure effects is due to Short Range
Correlated two-nucleon pairs ($NN$ SRC) which have large relative
momentum and small center-of-mass momentum.  SRC pairs account for
about 20\% of nucleons, almost all of the high momentum ($k>k_F\approx
250$ MeV/c) nucleons, and most of the kinetic energy in medium to
heavy nuclei
\cite{egiyan03,egiyan06,frankfurt93,fomin12,piasetzky06,subedi08,korover14}.
They are composed predominantly of neutron-proton $np$ correlated
pairs, even in heavy, neutron-rich, asymmetric nuclei
\cite{subedi08,hen14}.

Since the incident neutrino energy for each event is inferred from the detected final
state particles, it is important to include the effects of two nucleon
currents and SRC pairs when  analyzing  neutrino-nucleus reactions
\cite{fields13,fiorentini13b}.  

% Our knowledge of the composition and distribution of SRC comes
% predominantly from high energy electron scattering, where the electron
% scatters from a high-momentum ($k>k_F$) proton.  In one type of
% experiment, the scattered electron and the knocked out proton are
% detected and a large acceptance detector is placed to look for the
% correlated partners, if any, of the knocked-out proton, $A(e,e'p)$ and $A(e,e'pN)$
% \cite{subedi08,korover14,hen14}.  In this experiment, the two nucleons
% in the SRC are back-to-back in the initial nucleus.  Another type of experiment involved
% measuring the $^3$He$(e,e'pp)n$ reaction by scattering electrons from
% $^3$He and detecting the scattered electrons and the emitted protons
% in a $4\pi$ spectrometer \cite{niyazov03,baghdasaryan10}.  That
% experiment detected many events where the electron scattered from a
% single nucleon, and the residual spectator pair emerged back-to-back
% in the lab frame (in the final state).

\begin{figure} [htbp]
%\begin{center}
\includegraphics[width=1.5in]{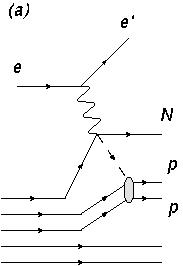}
\quad\includegraphics[width=1.5in]{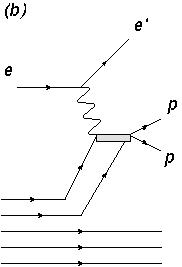}
\caption{\label{fig:res}  Pictorial diagram of electron
  induced two-proton knockout, $A(e,e'pp)$.   (a) The electron
  scatters from a first nucleon, emitting a pion.  The pion is absorbed by two
  nucleons, which are detected.  (b)  The electron
  scatters from a first nucleon, exciting it to a resonance, which
  deexcites via $\Delta N \rightarrow pp$.  The diagrams for the corresponding charged
  current neutrino interactions $A(\nu,\mu pp)$ would replace the
  incident electron with a neutrino, the outgoing electron with a
  muon, and the exchanged virtual photon with a $W$.}
%\end{center}
\end{figure}

The ArgoNeut large-volume liquid-argon time projection chamber (TPC) in the Main
Injector neutrino beam at Fermilab has detected 19 events with two
high-momentum ($k>k_F$) protons and no pions in the final state,
Ar$(\nu,\mu^-pp)$   \cite{acciardi14}.  Of these, 
four events are visually striking, with a long muon track and two protons back-to-back in the lab frame (hammer
events).  

The hammer events are remarkable because the two protons are
back-to-back ($\cos\theta_{pp}<-0.95$), high momentum ($p_p\approx500$
MeV/c), and have moderate to large missing transverse momentum
$p^T_{miss}\ge300$ MeV/c \cite{acciardi14}.  These are attributed to
resonance production on the struck nucleon, followed by pion emission
and absorption on the correlated pair (see Fig. \ref{fig:res}a).  The
authors claim that, ``The detection of back-to-back $pp$ pairs in the
lab frame can be seen as {\bf “snapshots” of the initial pair
  configuration} in the case of RES processes with no or low momentum
transfer to the pair (emphasis added).''  However, these events were
not described by a simulation using the standard NUWRO Monte Carlo
event generator \cite{niewczas16}, even though it included
quasielastic, resonant, inelastic, coherent pion production, and two
body current processes.

The much larger MicroBooNE liquid-argon TPC should detect far more of
these intriguing events.

In this paper we will develop two simple semi-classical models of these
hammer events.  The first model will describe pion production on a
nucleon followed by pion absorption on an $NN$ pair (see
Fig. \ref{fig:res}a) and the second model will describe resonance
excitation (primarily $\Delta(1232)$) followed by deexcitation via the
reaction $\Delta N \rightarrow pp$ (see Fig. \ref{fig:res}b).  

The pion-production model will provide the first semi-quantitative
explanation of the hammer events.  The results of this model will also show that,
contrary to the claims of Ref.~\cite{acciardi14}, the final state
distribution of $pp$ pairs is relatively insensitive to the details of
the ``initial pair configuration''.  On the positive side, these events can be used
to reconstruct the incident neutrino energy from the momentum of the
outgoing muon with or without the momenta of the two outgoing protons.

\noindent {\bf Methods:} We describe the reaction process of
Fig. \ref{fig:res}a by a sequential pion production and absorption model.  We
use the MAID-2000 \cite{drechsel99} parametrization to take advantage
of the well measured pion-electroproduction cross section.  While the
electron $(e,e'\pi)$ and charged-current neutrino $(\nu,\mu^-\pi)$
pion production cross sections differ in detail, both are
$\Delta(1232)$-dominated in this energy region and both are
transverse.  Therefore both processes should produce similar pion
momentum distributions.

We generate a nucleon with initial momentum according to the $^{12}$C
Argonne V18 momentum distribution \cite{wiringa14}, randomly sample
the initial electron energy from the MiniBoone neutrino energy
distribution \cite{aguilar09}, and uniformly generate the scattered
electron energy and angles and the emitted pion angles.  We then
calculate the cross section using MAID-2000.  We generate the center
of mass momentum of the $np$ pair ($\vec p_{CM} = \vec p_1 + \vec
p_2\thinspace$) using two models, the distribution of two uncorrelated
single nucleons using the $^{12}$C Argonne V18 momentum distribution
for each nucleon and the distribution of a correlated pair using a gaussian distribution
in each cartesian direction with $\sigma_x = \sigma_y=\sigma_z = 0.14$
GeV/c (as measured in Refs. \cite{tang03,shneor07} and calculated in
Ref. \cite{cda96}).  We then calculate the $\pi^+d$ absorption cross
section using the SAID-1998 \cite{said98} parametrization.  The final
momenta of the two protons are generated randomly in phase space
from the decay of the $np\pi^+\rightarrow pp$ system.

This model does not include the effects of final state interactions
(FSI), rescattering of the two final-state protons as they exit the
residual nucleus.  FSI will reduce the cross section significantly
(which is not relevant for this calculation) and somewhat smear the
momentum of the outgoing protons.

This model is insensitive to the initial relative momentum of the $np$
pair which absorbs the $\pi$ and therefore cannot distinguish between an SRC $np$ pair with
large relative momentum and a non-correlated neutron and proton with
small relative momentum.

In addition,  there was almost no difference in the 
distribution of the two outgoing protons whether we described the
center of mass momentum of the $np$ pair absorbing the pion as
the sum of two single-nucleon momenta or as the SRC
pair momentum distribution.

\begin{figure} [htbp]
%\begin{center}
\includegraphics[width=3in]{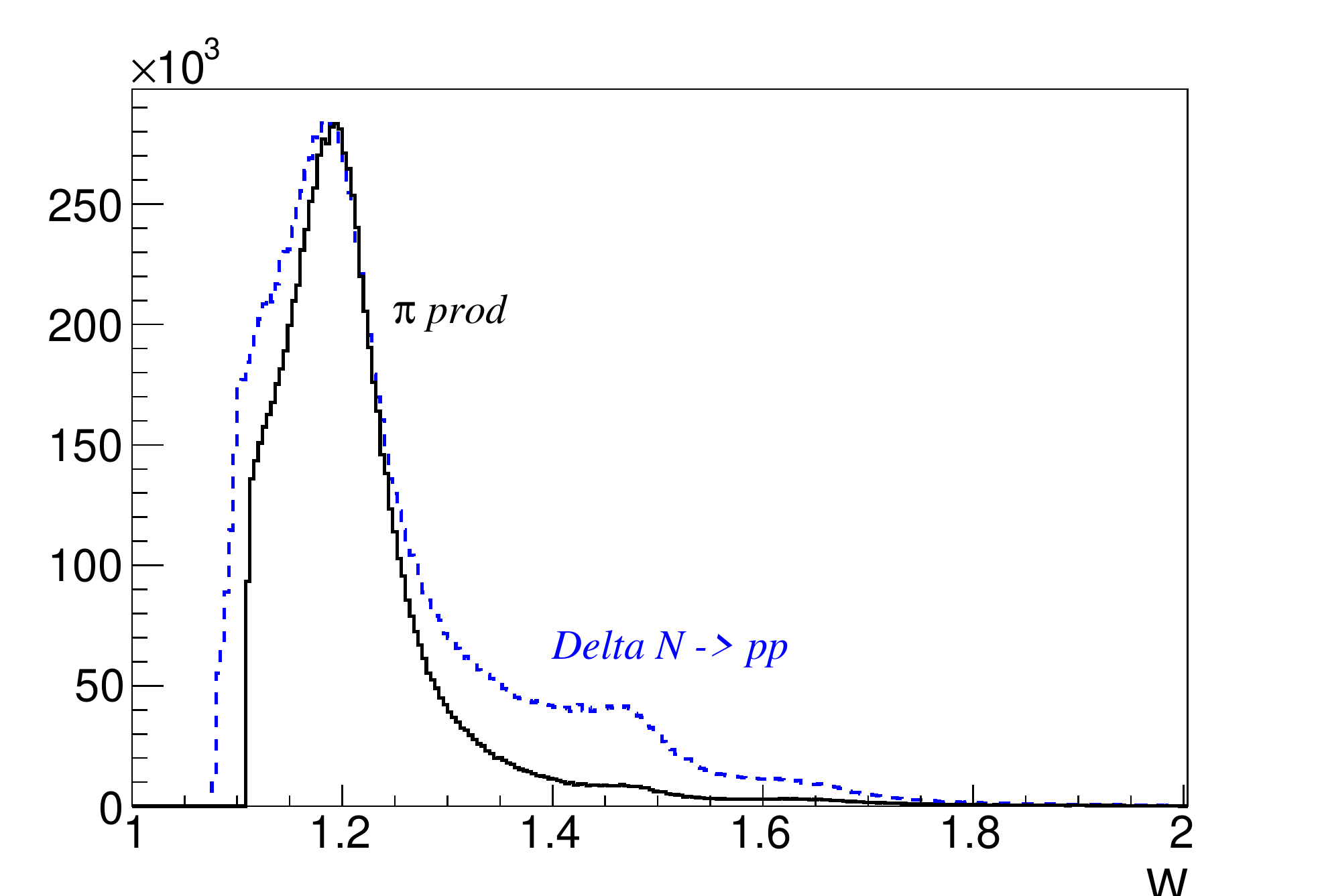}
\caption{\label{fig:W}  The invariant mass of the initial proton plus
  vector boson (e.g., the virtual photon) expected for hammer events
  in MicroBoone.  Black solid line: $\pi$
  production and reabsorption model; blue dashed line: $\Delta
  N\rightarrow pp$ model.}
%\end{center}
\end{figure}

\begin{figure} [htbp]
%\begin{center}
\includegraphics[width=3in]{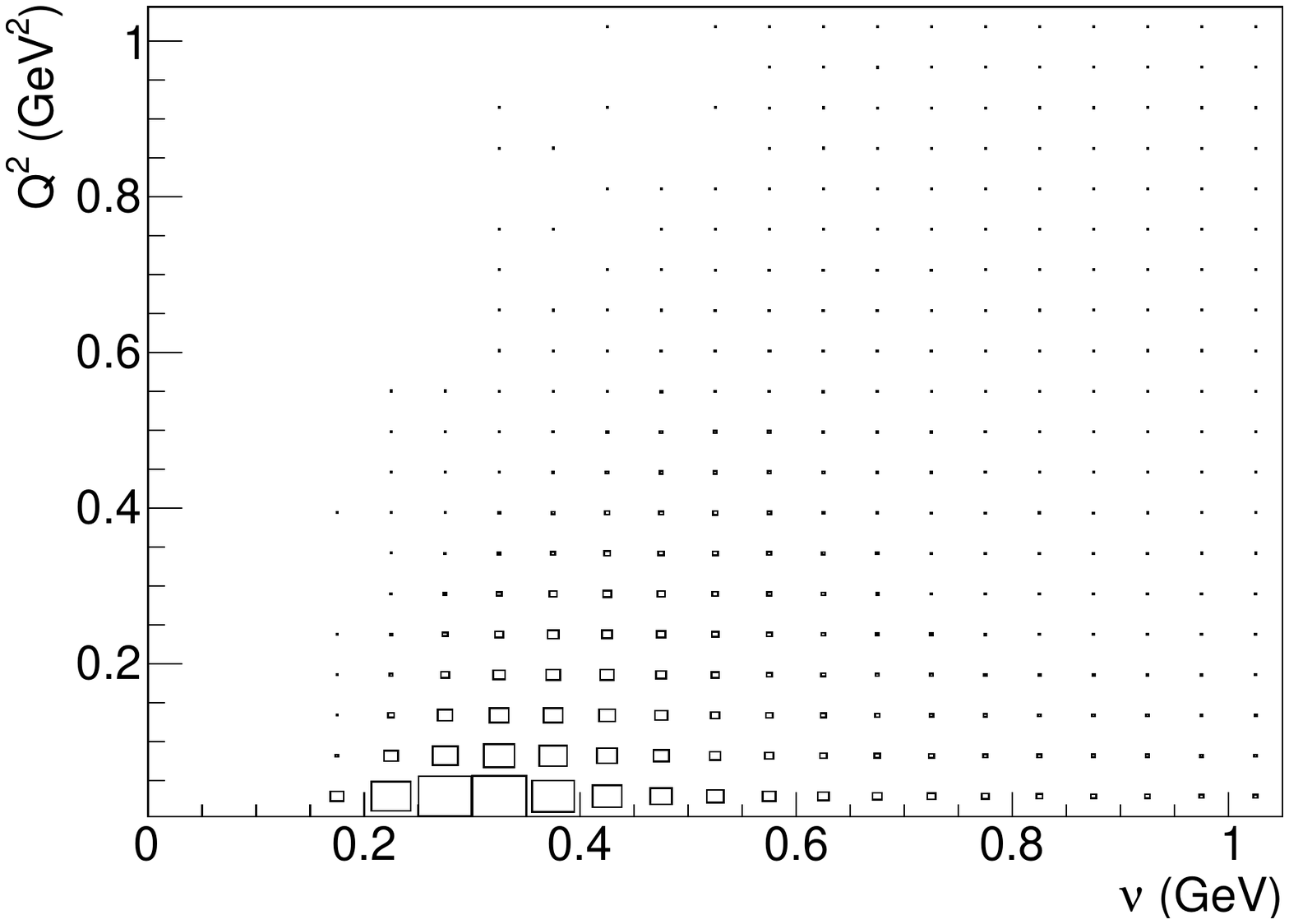}
\caption{\label{fig:q2nu} The momentum transfer squared, $Q^2$,
  plotted versus the energy transfer $\nu$ expected for hammer events
  in MicroBoone.}
%\end{center}
\end{figure}

% \begin{figure} [htbp]
% %\begin{center}
% \includegraphics[width=3in]{MomPion.pdf}
% \caption{\label{fig:mompi} The momentum of the pion exchanged between
%   the struck nucleon and the $np$ pair expected for hammer events
%   in MicroBoone.}
% %\end{center}
% \end{figure}

Because the MiniBoone incident neutrino energies are relatively low
(peaked at ~0.5 GeV with a tail extending to 2 GeV), the
reaction process is dominated by $\Delta(1232)$ production (see
Fig. \ref{fig:W}).  The pion-``deuteron'' absorption cross section
also peaks at the $\Delta$, further emphasizing the $\Delta$ peak.  The
momentum and energy transfer are small, with $Q^2$ starting
at zero and the energy transfer starting at the pion production
threshold (see Fig. \ref{fig:q2nu}).  Because the momentum transfer is
small, the momentum of the (unobserved) exchanged pion is also relatively small,
peaking at 0.2 GeV/c.
% (see Fig. \ref{fig:mompi}).

\begin{figure} [htbp]
%\begin{center}
\includegraphics[width=3.5in]{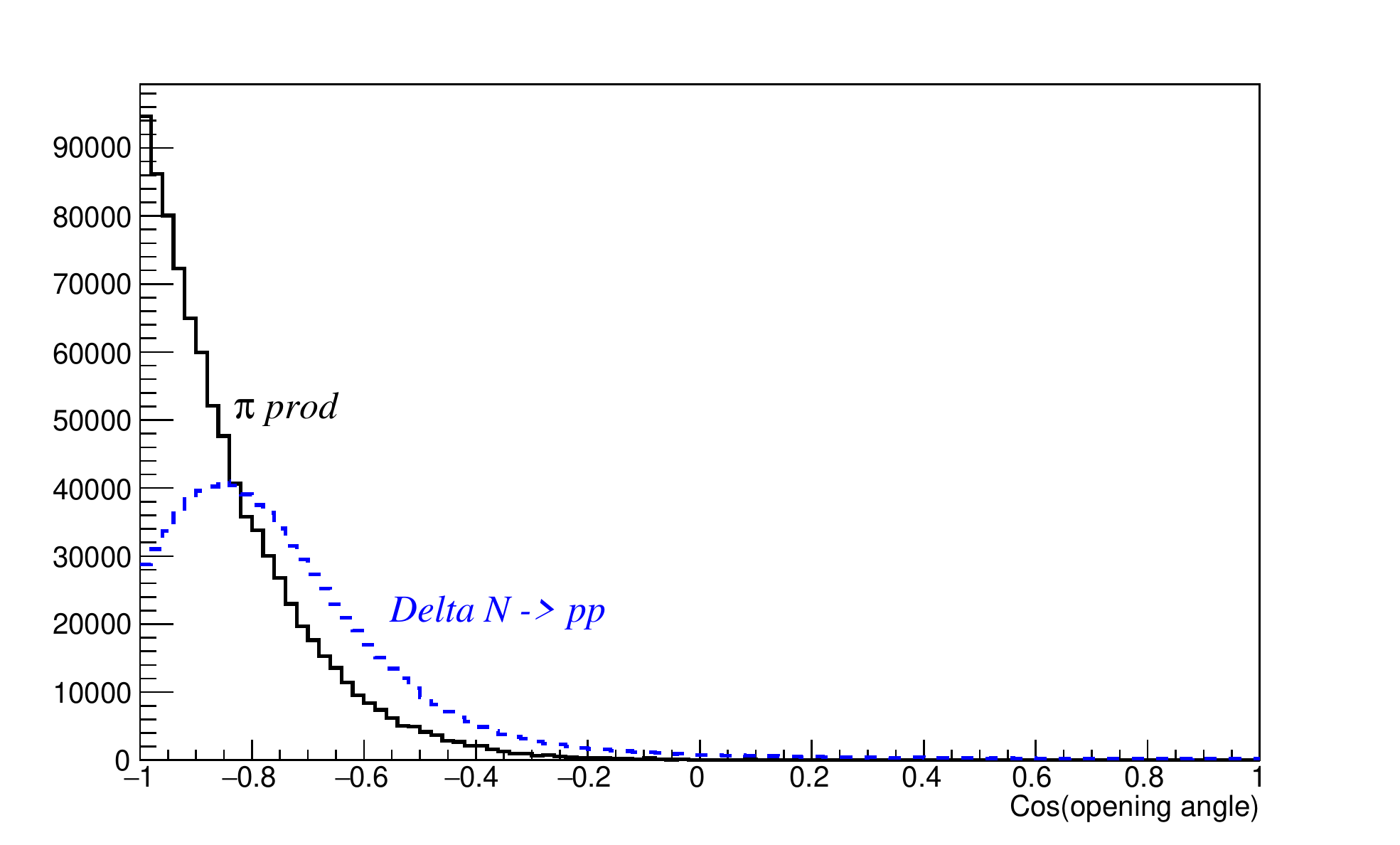}
\caption{\label{fig:openang} The cosine of the opening angle of the
  two final state protons in the lab system expected for hammer events
  in MicroBoone.  The black solid histogram
  corresponds to the pion production and reabsorption model and the
  blue dashed histogram corresponds to the
  $\Delta N\rightarrow pp$ model.  The opening angle distribution was
  almost identical for the two different center of mass momentum distributions
  of the $NN$ pair absorbing the $\pi$.}
%\end{center}
\end{figure}

\begin{figure} [htbp]
%\begin{center}
\includegraphics[width=3in]{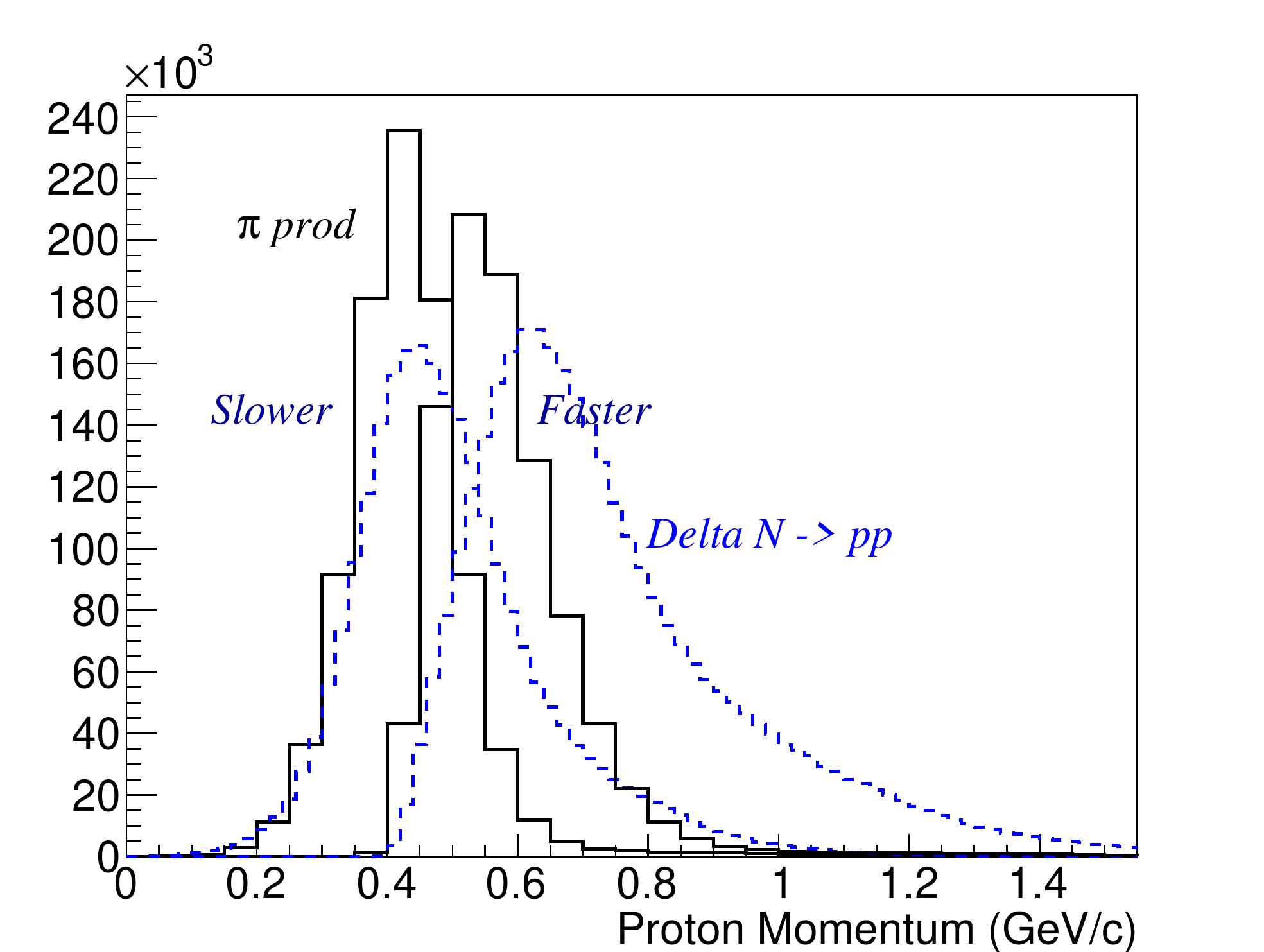}
\caption{\label{fig:momp} (color online) The momentum of the two protons in
  the lab frame, sorted into the slower proton and faster proton as expected for hammer events
  in MicroBoone.  The histograms peaked at $p_p\approx 0.4$
  GeV/c correspond to the slower proton and the histograms peaked
  at $p_p > 0.5$ to $0.6$
  GeV/c correspond to the faster proton
in the event.  The  solid black large-bin histogram corresponds to the pion
production and reabsorption model and the dashed blue small-bin histogram
corresponds to the $\Delta N\rightarrow pp$ model. }
%\end{center}
\end{figure}

\begin{figure} [htbp]
%\begin{center}
\includegraphics[width=3.5in]{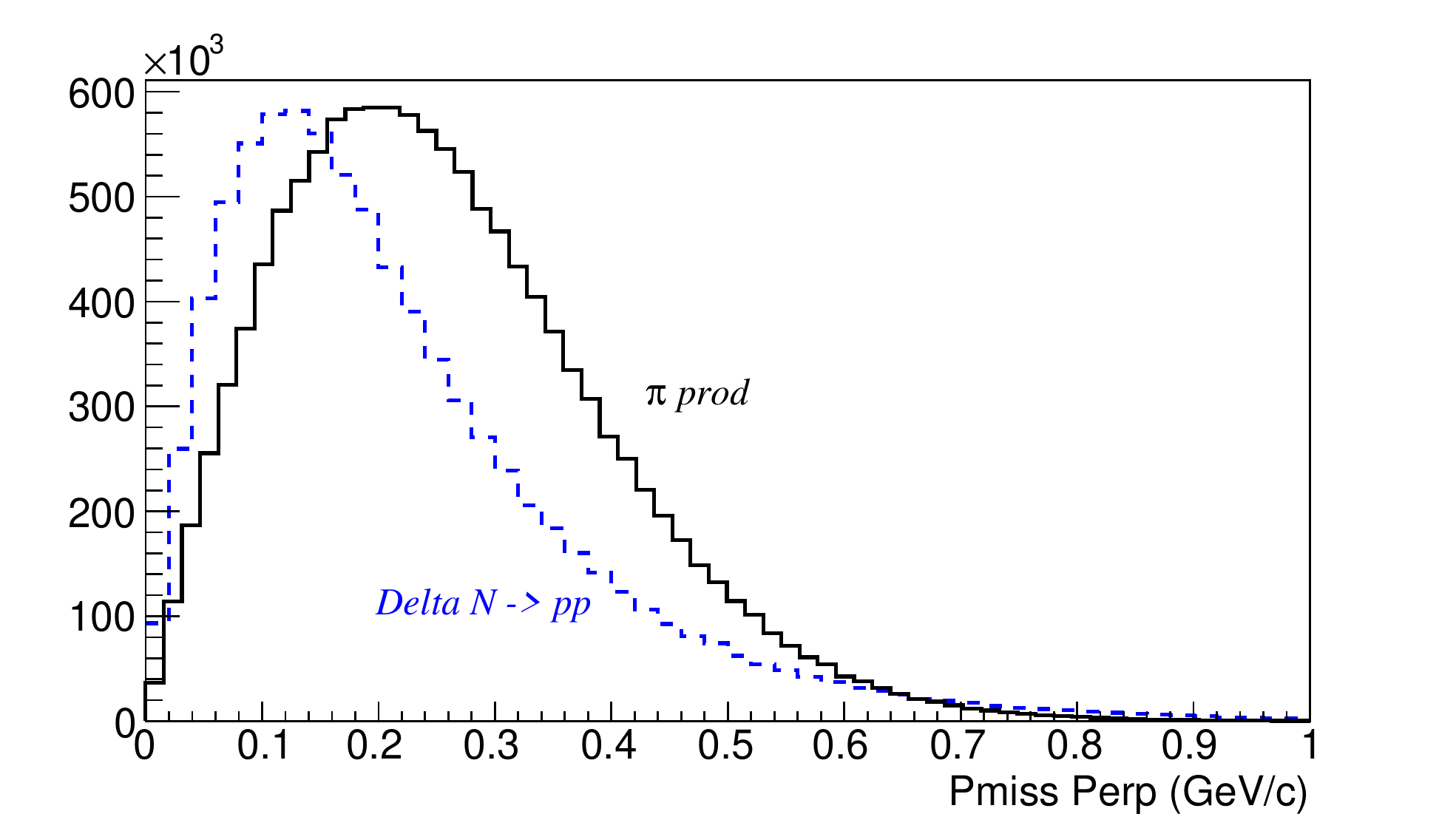}
\caption{\label{fig:pmperp} The perpendicular missing momentum of the
  muon plus
  two final state protons in the lab system expected for hammer events
  in MicroBoone.  The black solid histogram
  corresponds to the pion production and reabsorption model and the
  blue dashed histogram corresponds to the
  $\Delta N\rightarrow pp$ model.  }
%\end{center}
\end{figure}

The resulting opening angle and momentum distributions of the two
protons can be seen in Figs. \ref{fig:openang} and \ref{fig:momp}.
The opening angle is predominantly back-to-back and the proton
momentum distributions are peaked at about 0.5 GeV/c.  The
transverse missing momentum of the measured particles ($\vec
p_{miss}^T = \vec p_\mu^T + \vec p_{p_1}^T + \vec
p_{p_2}^T$) is peaked at about 0.2 GeV/c with a long tail
extending out to higher momenta (see Fig. \ref{fig:pmperp}).  

The opening angle and momentum distributions of the two protons are
consistent with the four observed hammer events
($\cos\theta_{pp}<-0.95$ and $p_p\approx500$ MeV/c). The missing
transverse momentum distribution is slightly smaller than the observed
$p_{miss}^T\ge0.3$ GeV/c.

We also calculated the expected results for an incident neutrino
energy of 4 GeV, comparable to the average energy of the ArgoNeut
neutrino beam.  The primary differences between these results and the
ones shown in the Figs. \ref{fig:W} and \ref{fig:q2nu}, are that the
reaction at these higher neutrino energies covers a much larger range
of energy and momentum transfer and a significantly wider range in
$W$.  However, the proton spectra
(Figs. \ref{fig:openang}--\ref{fig:pmperp}) are remarkably similar for
the two sets of incident enutrino energies.  The two
protons are predominantly back-to-back, with momenta peaked at about
500 MeV/c and the same $p_{miss}^T$ distribution.

We also compared these distributions to those from the second reaction
model, where the incident neutrino scatters from a nucleon, exciting
it to a resonance ($N^*$ or $\Delta$), which then deexcites by
colliding with a second nucleon, e.g., $\Delta N\rightarrow pp$.  This
model generates a nucleon with initial momentum according to the
$^{12}$C Argonne V18 momentum distribution \cite{wiringa14}.  It
generates the initial electron energy by randomly sampling from the
MiniBoone neutrino energy distribution \cite{aguilar09}.  It then
generates the scattered electron energy and angles.  It assumes that
the inclusive $\Delta$ production cross section has the same invariant
mass ($W$) dependence as the $eN\rightarrow e \pi N$ cross section
averaged over all outgoing pion momenta.  It then generates a second
nucleon randomly, and calculates the two-nucleon $\Delta N\rightarrow
pp$ distribution randomly by phase space.

The invariant mass distribution in this model is still peaked at the
$\Delta(1232)$ mass, although less strongly peaked than the pion
production and reabsorption model (Fig. \ref{fig:W}).
The protons from this model are significantly less back-to-back
(Fig. \ref{fig:openang}), have higher momentum (Fig. \ref{fig:momp}),
and have less missing transverse momentum (Fig. \ref{fig:pmperp}) than
the pion production and reabsorption model.  This the results of this model agree less
well with the four observed hammer events.

There are four possible reaction channels in the pion production and
reabsorption model leading to two back-to-back protons in
the final state, three for neutrinos and one
for anti-neutrinos:
\begin{align}
\nu &\rightarrow \mu^-W^+;\quad W^+n\rightarrow \pi^0\mathbf{p}; \quad \pi^0
pp\rightarrow \mathbf{pp} \\
\nu &\rightarrow \mu^-W^+;\quad W^+n\rightarrow \pi^+\mathbf{n}; \quad \pi^+
np\rightarrow \mathbf{pp} \\
\nu &\rightarrow \mu^-W^+;\quad W^+p\rightarrow \pi^+\mathbf{p}; \quad \pi^+
np\rightarrow \mathbf{pp} \\
\bar\nu &\rightarrow \mu^+W^-;\quad W^-p\rightarrow \pi^0\mathbf{n}; \quad \pi^0
pp\rightarrow \mathbf{pp} 
\end{align}
where the nucleons in boldface type are in the final state and can be
detected.  There is only one reaction channel each for $\nu$ and
$\bar\nu$ that lead to two back-to-back protons plus a neutron in the
final state.  There are two more $\nu$ reaction
channels that lead to two back-to-back protons plus a third proton in the
final state.   There is one reaction channel for the $\Delta N
\rightarrow pp$ reaction,
\[
\nu \rightarrow \mu^-W^+; \quad W^+N\rightarrow \Delta;\quad \Delta N
\rightarrow \mathbf{pp}\quad .
\]

The cross section for $\pi$ absorption on an isospin $T=1$ $NN$ pair (e.g., $pp$)
is about  ten times smaller than that for $\pi$ absorption on
a $T=0$ $NN$ pair \cite{ericson1988pions}.  Therefore reaction (1) will be suppressed by a
factor of ten relative to reaction  (3) which leads to the same final
state.  Reaction (2), which produces the characteristic
hammer signature, should be the same size as reaction (3), which
produces the hammer signature plus another proton.  

Therefore, in this model, there should be about equal numbers of
hammer events with an extra neutron as with an extra proton.  Reaction
channel (4) should also be ten times smaller than reaction channel (3)
so that there should be ten times fewer anti-neutrino events than
neutrino events.
%However, measuring this requires measuring the charge of the outgoing $\mu$.

\begin{figure} [htbp]
\begin{center}
\includegraphics[width=3.2in]{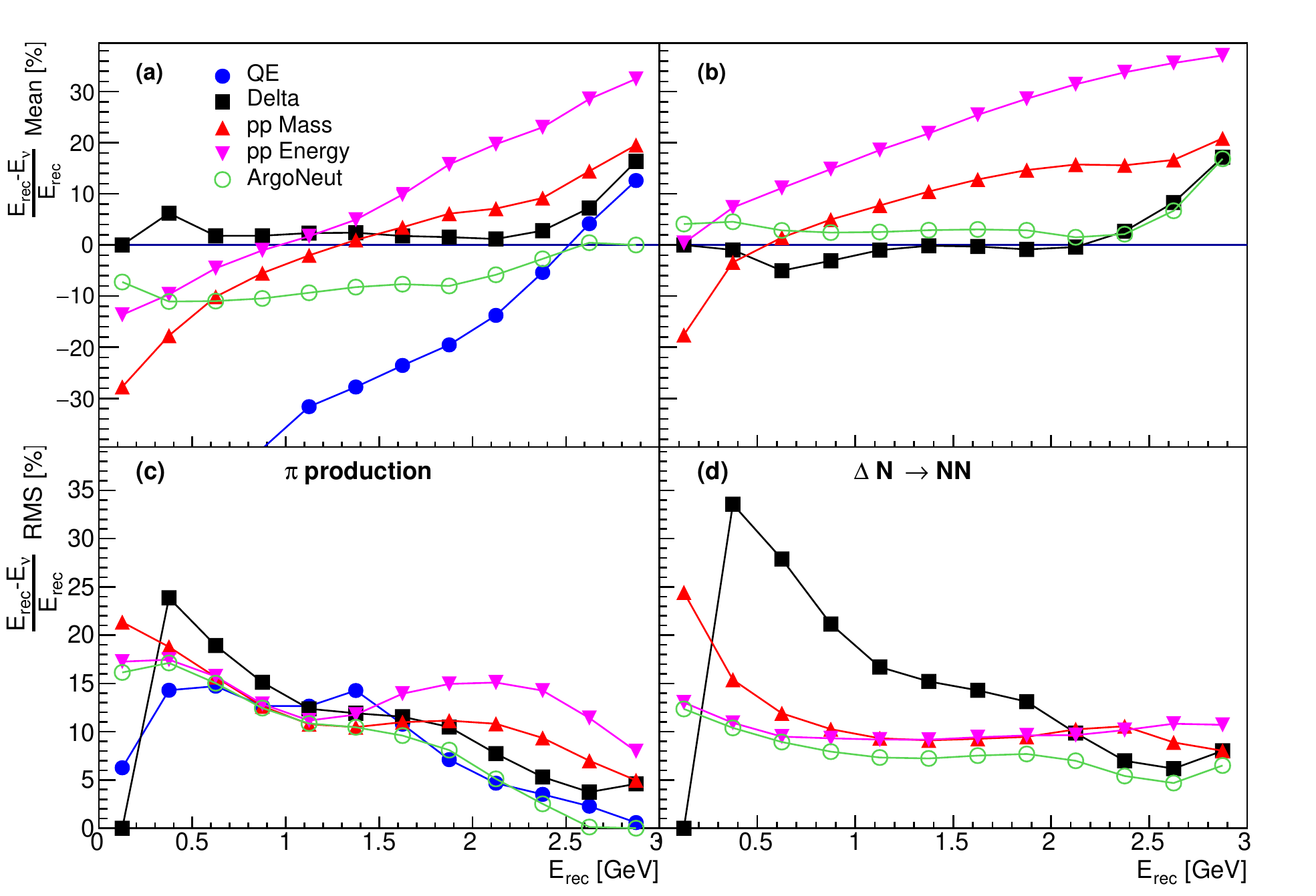}
\caption{\label{fig:recon}  The fractional error (top) and standard deviation
    (bottom) of the distribution of the reconstructed
  beam energy $(E_{rec} - E_\nu)/E_{rec}$ vs the reconstructed
  beam energy  for four ways to calculate the invariant mass of the struck
  nucleon as described in the text.  (left) for the $\pi$ production and reabsorption model
  calculated and (right) for the $N\Delta\rightarrow pp$ model.  The
  labels QE, Delta, pp Mass, and pp Energy refer to the four models of
the invariant mass of the struck nucleon described in the text.  The
label ArgoNeut refers to $E_\nu = E_\mu + T_{p1} + T_{p2} + T_{A-2} +
30$ MeV as used in Ref. \cite{acciardi14}.}
\end{center}
\end{figure}

% \begin{figure} [htbp]
% \begin{center}
% \includegraphics[width=3in]{NuEnergyErrorND.pdf}
% \includegraphics[width=3in]{NuEnergyWidthND.pdf}
% \caption{\label{fig:reconND}  The fractional error (top) and standard deviation
%     (bottom) of the distribution of the reconstructed
%   beam energy $(E_{rec} - E_\nu)/E_{rec}$ vs the reconstructed
%   beam energy  for the $N\Delta\rightarrow pp$ model calculated using
%   three different methods as described in the text.}
% \end{center}
% \end{figure}

ArgoNeut used a total energy method to reconstruct the
incident neutrino energy \cite{acciardi14} where $E_\nu = E_\mu +
T_{p1} + T_{p2} + T_{A-2} + 30$ MeV.   

Because these hammer events appear to all be due to the pion
production and reabsorption model, we can use a different algorithm to determine the incident neutrino energy
of the reaction.  The reconstructed incident neutrino energy, $E_{rec}$, depends on the 
energy, $E'$, and angle, $\theta$ of the outgoing lepton and on the invariant mass of
the struck nucleon plus transferred vector boson, $m'^2=(p^\mu_N +
p^\mu_B)^2$ (where $p^\mu_N$ and $p^\mu_B$ are the four vectors of the
struck nucleon and the transferred vector boson respectively): 
\begin{equation}
E_{rec} = \frac{m'^2 - m_N^2 + 2m_NE'}{2(m_N-E'(1-\cos\theta))}
\end{equation}
We can model the unknown invariant mass of the struck nucleon in
several ways (see Fig. \ref{fig:recon}):
\begin{itemize}
\item (``QE'') assume that the neutrino scattered quasielastically from a
  nucleon ($m'=m_N$);
\item (``Delta'') assume that the neutrino scattered from a nucleon, exciting it
  to a $\Delta$ ($m'=m_\Delta = 1.232$ GeV/c$^2$);
\item (``pp Mass'') assume that the extra invariant mass of the two protons equals the
  excitation energy of the struck nucleon ($m'=m_N + (m_{pp} -
  2m_N)$); and
\item (``pp Energy'') assume that the kinetic energy of the two protons equals the
  excitation energy of the struck nucleon ($m'=m_N + T_{p_1} +
  T_{p_2}$).
\end{itemize}

Fig. \ref{fig:recon} shows the
accuracy (mean) and precision ($\sigma$) of the reconstructed neutrino
energies for both the $\pi$ production and reabsorption model and the
$\Delta N\rightarrow pp$ model.  The accuracy is defined as the mean
of the $(E_{rec} - E_\nu)/E_{rec}$ distribution and the precision is
defined as the RMS of the $(E_{rec} - E_\nu)/E_{rec}$ distribution.

In the pion production and reabsorption model, assuming that the
struck nucleon is excited to a $\Delta$ gives the most accurate
reconstruction of the incident neutrino energy over the entire range
of energies.  Using the two-proton kinetic energy or invariant mass to
estimate the excitation energy of the struck nucleon is less accurate.
The ArgoNeut total energy method is also less accurate, due to the
presence of undetected energetic particles.
Assuming that the reaction was quasielastic fails completely.  

In the $\Delta N\rightarrow pp$ model, the ArgoNeut total energy
method works the best because there are no energetic undetected
particles.  Assuming the struck nucleon is excited to a $\Delta$ is
equally accurate but significantly less precise.

% Assuming that the struck nucleon is excited to a $\Delta$ lets us
% reconstruct the initial neutrino energy to about 30\% at 0.5 GeV,
% improving to $\approx$15\% at higher energies.

\noindent {\bf Summary:}
The ArgoNeut liquid Argon Time Projection Chamber observed four
charged-current neutrino-argon scattering events with two protons
back-to-back in the final state (“hammer” events).  These events were
attributed to resonance production on the struck nucleon, followed by
pion emission and absorption on a correlated pair \cite{acciardi14}.  These events were
not described by the NUWRO Monte Carlo event generator \cite{niewczas16}.

We modeled these
hammer events with a semi-classical model where the lepton scatters from a
moving nucleon, causing it to emit a $\pi$.  The $\pi$ is then absorbed by
two nucleons ($NN$).  
%We used the MiniBoone incident neutrino energy
%distribution, the well known pion-electroproduction cross section, and
%the measured pion-deutron absorption cross section.  We used two
%models of the $NN$ center of mass momentum distribution, the measured
%Gaussian distribution of a short range correlated pair and the sum of
%two single particle momentum distributions.  
This pion production and
reabsorption process results in events with two back-to-back protons
each with momentum of about 500 MeV/c and moderate transverse missing
momentum, very similar to that of the observed ArgoNeut events.  The
results of this model are completely insensitive to the relative
momentum of the $NN$ pair and to the choice of its center of mass
momentum distribution.  This model predicts that a third nucleon is
emitted from the nucleus and that about half the time this third
nucleon is an easily detectable proton.  In this model, the incident
neutrino energy can be reconstructed accurately using just
the outgoing lepton momentum and angle (for the relatively low MicroBooNE
neutrino energies).  This energy reconstruction is significantly
better than the standard ArgoNeut total energy method.

We also modeled nucleon excitation, followed by deexcitation via
$\Delta p\rightarrow pp$.  This process results in two protons that
are less similar to the observed events.  We should be able to
decisively distinguish between the two models by the fraction of
hammer events with a third emitted proton.

We conclude that ArgoNeut hammer events can be described by a simple pion production
and reabsorption model. These events can be used to determine the
incident neutrino energy, but cannot teach us anything significant
about short range correlated $NN$ pairs.  We suggest
that this reaction channel could be used for neutrino oscillation
experiments to complement other channels with higher statistics but
greater uncertainty in the incident neutrino energy.

\begin{acknowledgments}

  We thank O. Palamara, F. Cavanna, and Sam Zeller for many fruitful
  discussions. This work was partially supported by the US Department
  of Energy under grants DE-FG02-97ER-41014, DE-FG02-96ER-40960,
  DE-FG02-01ER-41172 and by the Israel Science Foundation.
\end{acknowledgments}

\bibliography{eep}

\begin{thebibliography}{21}
\expandafter\ifx\csname natexlab\endcsname\relax\def\natexlab#1{#1}\fi
\expandafter\ifx\csname bibnamefont\endcsname\relax
  \def\bibnamefont#1{#1}\fi
\expandafter\ifx\csname bibfnamefont\endcsname\relax
  \def\bibfnamefont#1{#1}\fi
\expandafter\ifx\csname citenamefont\endcsname\relax
  \def\citenamefont#1{#1}\fi
\expandafter\ifx\csname url\endcsname\relax
  \def\url#1{\texttt{#1}}\fi
\expandafter\ifx\csname urlprefix\endcsname\relax\def\urlprefix{URL }\fi
\providecommand{\bibinfo}[2]{#2}
\providecommand{\eprint}[2][]{\url{#2}}

\bibitem[{\citenamefont{Mosel}(2016)}]{mosel16}
\bibinfo{author}{\bibfnamefont{U.}~\bibnamefont{Mosel}}, \bibinfo{journal}{Adv.
  Nucl. Part. Sci.} \textbf{\bibinfo{volume}{66}} (\bibinfo{year}{2016}),
  \eprint{1602.00696}.

\bibitem[{\citenamefont{Egiyan et~al.}(2003)}]{egiyan03}
\bibinfo{author}{\bibfnamefont{K.}~\bibnamefont{Egiyan}} \bibnamefont{et~al.}
  (\bibinfo{collaboration}{CLAS Collaboration}), \bibinfo{journal}{Phys. Rev.
  C} \textbf{\bibinfo{volume}{68}}, \bibinfo{pages}{014313}
  (\bibinfo{year}{2003}).

\bibitem[{\citenamefont{Egiyan et~al.}(2006)}]{egiyan06}
\bibinfo{author}{\bibfnamefont{K.}~\bibnamefont{Egiyan}} \bibnamefont{et~al.}
  (\bibinfo{collaboration}{CLAS Collaboration}), \bibinfo{journal}{Phys. Rev.
  Lett.} \textbf{\bibinfo{volume}{96}}, \bibinfo{pages}{082501}
  (\bibinfo{year}{2006}).

\bibitem[{\citenamefont{Frankfurt et~al.}(1993)\citenamefont{Frankfurt,
  Strikman, Day, and Sargsyan}}]{frankfurt93}
\bibinfo{author}{\bibfnamefont{L.}~\bibnamefont{Frankfurt}},
  \bibinfo{author}{\bibfnamefont{M.}~\bibnamefont{Strikman}},
  \bibinfo{author}{\bibfnamefont{D.}~\bibnamefont{Day}}, \bibnamefont{and}
  \bibinfo{author}{\bibfnamefont{M.}~\bibnamefont{Sargsyan}},
  \bibinfo{journal}{Phys. Rev. C} \textbf{\bibinfo{volume}{48}},
  \bibinfo{pages}{2451} (\bibinfo{year}{1993}).

\bibitem[{\citenamefont{Fomin et~al.}(2012)}]{fomin12}
\bibinfo{author}{\bibfnamefont{N.}~\bibnamefont{Fomin}} \bibnamefont{et~al.},
  \bibinfo{journal}{Phys. Rev. Lett.} \textbf{\bibinfo{volume}{108}},
  \bibinfo{pages}{092502} (\bibinfo{year}{2012}).

\bibitem[{\citenamefont{Piasetzky et~al.}(2006)\citenamefont{Piasetzky,
  Sargsian, Frankfurt, Strikman, and Watson}}]{piasetzky06}
\bibinfo{author}{\bibfnamefont{E.}~\bibnamefont{Piasetzky}},
  \bibinfo{author}{\bibfnamefont{M.}~\bibnamefont{Sargsian}},
  \bibinfo{author}{\bibfnamefont{L.}~\bibnamefont{Frankfurt}},
  \bibinfo{author}{\bibfnamefont{M.}~\bibnamefont{Strikman}}, \bibnamefont{and}
  \bibinfo{author}{\bibfnamefont{J.~W.} \bibnamefont{Watson}},
  \bibinfo{journal}{Phys. Rev. Lett.} \textbf{\bibinfo{volume}{97}},
  \bibinfo{pages}{162504} (\bibinfo{year}{2006}).

\bibitem[{\citenamefont{Subedi et~al.}(2008)}]{subedi08}
\bibinfo{author}{\bibfnamefont{R.}~\bibnamefont{Subedi}} \bibnamefont{et~al.},
  \bibinfo{journal}{Science} \textbf{\bibinfo{volume}{320}},
  \bibinfo{pages}{1476} (\bibinfo{year}{2008}).

\bibitem[{\citenamefont{Korover et~al.}(2014)\citenamefont{Korover, Muangma,
  Hen et~al.}}]{korover14}
\bibinfo{author}{\bibfnamefont{I.}~\bibnamefont{Korover}},
  \bibinfo{author}{\bibfnamefont{N.}~\bibnamefont{Muangma}},
  \bibinfo{author}{\bibfnamefont{O.}~\bibnamefont{Hen}}, \bibnamefont{et~al.},
  \bibinfo{journal}{Phys.Rev.Lett.} \textbf{\bibinfo{volume}{113}},
  \bibinfo{pages}{022501} (\bibinfo{year}{2014}), \eprint{1401.6138}.

\bibitem[{\citenamefont{Hen et~al.}(2014)}]{hen14}
\bibinfo{author}{\bibfnamefont{O.}~\bibnamefont{Hen}} \bibnamefont{et~al.}
  (\bibinfo{collaboration}{CLAS Collaboration}), \bibinfo{journal}{Science}
  \textbf{\bibinfo{volume}{346}}, \bibinfo{pages}{614} (\bibinfo{year}{2014}).

\bibitem[{\citenamefont{Fields et~al.}(2013)}]{fields13}
\bibinfo{author}{\bibfnamefont{L.}~\bibnamefont{Fields}} \bibnamefont{et~al.}
  (\bibinfo{collaboration}{MINERvA Collaboration}), \bibinfo{journal}{Phys.
  Rev. Lett.} \textbf{\bibinfo{volume}{111}}, \bibinfo{pages}{022501}
  (\bibinfo{year}{2013}),
  \urlprefix\url{http://link.aps.org/doi/10.1103/PhysRevLett.111.022501}.

\bibitem[{\citenamefont{Fiorentini et~al.}(2013)}]{fiorentini13b}
\bibinfo{author}{\bibfnamefont{G.~A.} \bibnamefont{Fiorentini}}
  \bibnamefont{et~al.} (\bibinfo{collaboration}{MINERvA Collaboration}),
  \bibinfo{journal}{Phys. Rev. Lett.} \textbf{\bibinfo{volume}{111}},
  \bibinfo{pages}{022502} (\bibinfo{year}{2013}).

\bibitem[{\citenamefont{Acciarri et~al.}(2014)}]{acciardi14}
\bibinfo{author}{\bibfnamefont{R.}~\bibnamefont{Acciarri}}
  \bibnamefont{et~al.}, \bibinfo{journal}{Phys. Rev. D}
  \textbf{\bibinfo{volume}{90}}, \bibinfo{pages}{012008}
  (\bibinfo{year}{2014}),
  \urlprefix\url{http://link.aps.org/doi/10.1103/PhysRevD.90.012008}.

\bibitem[{\citenamefont{Niewczas and Sobczyk}(2016)}]{niewczas16}
\bibinfo{author}{\bibfnamefont{K.}~\bibnamefont{Niewczas}} \bibnamefont{and}
  \bibinfo{author}{\bibfnamefont{J.~T.} \bibnamefont{Sobczyk}},
  \bibinfo{journal}{Phys. Rev. C} \textbf{\bibinfo{volume}{93}},
  \bibinfo{pages}{035502} (\bibinfo{year}{2016}),
  \urlprefix\url{http://link.aps.org/doi/10.1103/PhysRevC.93.035502}.

\bibitem[{\citenamefont{Drechsel et~al.}(1999)\citenamefont{Drechsel, Hanstein,
  Kamalov, and Tiator}}]{drechsel99}
\bibinfo{author}{\bibfnamefont{D.}~\bibnamefont{Drechsel}},
  \bibinfo{author}{\bibfnamefont{O.}~\bibnamefont{Hanstein}},
  \bibinfo{author}{\bibfnamefont{S.}~\bibnamefont{Kamalov}}, \bibnamefont{and}
  \bibinfo{author}{\bibfnamefont{L.}~\bibnamefont{Tiator}},
  \bibinfo{journal}{Nuclear Physics A} \textbf{\bibinfo{volume}{645}},
  \bibinfo{pages}{145 } (\bibinfo{year}{1999}), ISSN \bibinfo{issn}{0375-9474},
  \urlprefix\url{http://www.sciencedirect.com/science/article/pii/S03759474980%
05727}.

\bibitem[{\citenamefont{Wiringa et~al.}(2014)\citenamefont{Wiringa, Schiavilla,
  Pieper, and Carlson}}]{wiringa14}
\bibinfo{author}{\bibfnamefont{R.~B.} \bibnamefont{Wiringa}},
  \bibinfo{author}{\bibfnamefont{R.}~\bibnamefont{Schiavilla}},
  \bibinfo{author}{\bibfnamefont{S.~C.} \bibnamefont{Pieper}},
  \bibnamefont{and} \bibinfo{author}{\bibfnamefont{J.}~\bibnamefont{Carlson}},
  \bibinfo{journal}{Phys. Rev. C} \textbf{\bibinfo{volume}{89}},
  \bibinfo{pages}{024305} (\bibinfo{year}{2014}).

\bibitem[{\citenamefont{Aguilar-Arevalo et~al.}(2009)}]{aguilar09}
\bibinfo{author}{\bibfnamefont{A.~A.} \bibnamefont{Aguilar-Arevalo}}
  \bibnamefont{et~al.} (\bibinfo{collaboration}{MiniBooNE Collaboration}),
  \bibinfo{journal}{Phys. Rev. D} \textbf{\bibinfo{volume}{79}},
  \bibinfo{pages}{072002} (\bibinfo{year}{2009}),
  \urlprefix\url{http://link.aps.org/doi/10.1103/PhysRevD.79.072002}.

\bibitem[{\citenamefont{Tang et~al.}(2003)}]{tang03}
\bibinfo{author}{\bibfnamefont{A.}~\bibnamefont{Tang}} \bibnamefont{et~al.},
  \bibinfo{journal}{Phys. Rev. Lett.} \textbf{\bibinfo{volume}{90}},
  \bibinfo{pages}{042301} (\bibinfo{year}{2003}).

\bibitem[{\citenamefont{Shneor et~al.}(2007)}]{shneor07}
\bibinfo{author}{\bibfnamefont{R.}~\bibnamefont{Shneor}} \bibnamefont{et~al.},
  \bibinfo{journal}{Phys. Rev. Lett.} \textbf{\bibinfo{volume}{99}},
  \bibinfo{eid}{072501} (\bibinfo{year}{2007}).

\bibitem[{\citenamefont{Ciofi~degli Atti and Simula}(1996)}]{cda96}
\bibinfo{author}{\bibfnamefont{C.}~\bibnamefont{Ciofi~degli Atti}}
  \bibnamefont{and} \bibinfo{author}{\bibfnamefont{S.}~\bibnamefont{Simula}},
  \bibinfo{journal}{Phys. Rev. C} \textbf{\bibinfo{volume}{53}},
  \bibinfo{pages}{1689} (\bibinfo{year}{1996}),
  \urlprefix\url{http://link.aps.org/doi/10.1103/PhysRevC.53.1689}.

\bibitem[{\citenamefont{Oh et~al.}(1997)\citenamefont{Oh, Arndt, Strakovsky,
  and Workman}}]{said98}
\bibinfo{author}{\bibfnamefont{C.~H.} \bibnamefont{Oh}},
  \bibinfo{author}{\bibfnamefont{R.~A.} \bibnamefont{Arndt}},
  \bibinfo{author}{\bibfnamefont{I.~I.} \bibnamefont{Strakovsky}},
  \bibnamefont{and} \bibinfo{author}{\bibfnamefont{R.~L.}
  \bibnamefont{Workman}}, \bibinfo{journal}{Phys. Rev. C}
  \textbf{\bibinfo{volume}{56}}, \bibinfo{pages}{635} (\bibinfo{year}{1997}),
  \urlprefix\url{http://link.aps.org/doi/10.1103/PhysRevC.56.635}.

\bibitem[{\citenamefont{Ericson and Weise}(1988)}]{ericson1988pions}
\bibinfo{author}{\bibfnamefont{T.}~\bibnamefont{Ericson}} \bibnamefont{and}
  \bibinfo{author}{\bibfnamefont{W.}~\bibnamefont{Weise}},
  \emph{\bibinfo{title}{Pions and nuclei}}, Oxford Science Publications
  (\bibinfo{publisher}{Clarendon Press}, \bibinfo{year}{1988}), ISBN
  \bibinfo{isbn}{9780198520085},
  \urlprefix\url{https://books.google.com/books?id=v099AAAAIAAJ}.

\end{thebibliography}

\end{document}